\documentclass[copyright,creativecommons]{eptcs}
\providecommand{\event}{TiCSA 2023} 

\usepackage{iftex}

\ifpdf
  \usepackage{underscore}         
  \usepackage[T1]{fontenc}        
\else
  \usepackage{breakurl}           
\fi

\usepackage[english]{babel}
\usepackage[utf8]{inputenc}
\usepackage{microtype}
\usepackage{graphicx}
\usepackage{wrapfig}
\usepackage{newlfont}
\usepackage{amssymb}
\usepackage{mathtools}
\usepackage[bbgreekl]{mathbbol}
\usepackage{latexsym}
\usepackage{amsthm}
\usepackage{inconsolata}
\usepackage{listings}
\usepackage[x11names,dvipsnames,table]{xcolor}
\usepackage{tikz}
\usetikzlibrary{calc}
\usepackage{adjustbox}

\usepackage{amsmath}
\usepackage[capitalise,noabbrev]{cleveref}
\usepackage{xspace}

\newcommand\YAMLcolonstyle{\color{BrickRed}\ttfamily}
\newcommand\YAMLkeystyle{\color{black}\ttfamily}
\newcommand\YAMLvaluestyle{\color{blue}\ttfamily}

\makeatletter

\newcommand\language@yaml{yaml}

\expandafter\expandafter\expandafter\lstdefinelanguage
\expandafter{\language@yaml}
{
  keywords={workers,controller,strategy,followup,invalidate,topology_tolerance},
  keywordstyle=\ttfamily\color{RoyalBlue},
  basicstyle=\YAMLkeystyle,                                 
  tabsize=1,
  sensitive=false,
  comment=[l]{\#},
  morecomment=[s]{/*}{*/},
  commentstyle=\color{gray}\ttfamily,
  stringstyle=\YAMLvaluestyle\ttfamily,
  moredelim=[l][\color{orange}]{\&},
  moredelim=**[is][\YAMLvaluestyle]{~}{~},
  morestring=[b]',
  morestring=[b]",
  literate =    {---}{{\ProcessThreeDashes}}3
                {>}{{\textcolor{red}\textgreater}}1     
                {|}{{\textcolor{red}\textbar}}1 
                {\ -\ }{{\mdseries\ -\ }}3
                {:}{{{\YAMLcolonstyle{:}}}}1
}

\lst@AddToHook{EveryLine}{\ifx\lst@language\language@yaml\YAMLkeystyle\fi}
\makeatother

\newcommand\JSONnumbervaluestyle{\color{blue}}
\newcommand\JSONstringvaluestyle{\color{red}}

\newif\ifcolonfoundonthisline

\makeatletter

\lstdefinestyle{json}
{
  showstringspaces    = false,
  keywords            = {false,true},
  alsoletter          = 0123456789.,
  morestring          = [s]{"}{"},
  stringstyle         = \ifcolonfoundonthisline\JSONstringvaluestyle\fi,
  MoreSelectCharTable =%
    \lst@DefSaveDef{`:}\colon@json{\processColon@json},
  basicstyle          = \ttfamily,
  keywordstyle        = \ttfamily\bfseries,
}

\newcommand\processColon@json{%
  \colon@json%
  \ifnum\lst@mode=\lst@Pmode%
    \global\colonfoundonthislinetrue%
  \fi
}

\lst@AddToHook{Output}{%
  \ifcolonfoundonthisline%
    \ifnum\lst@mode=\lst@Pmode%
      \def\lst@thestyle{\JSONnumbervaluestyle}%
    \fi
  \fi
  \lsthk@DetectKeywords%
}

\lst@AddToHook{EOL}%
  {\global\colonfoundonthislinefalse}

\makeatother

\newcommand\ProcessThreeDashes{\llap{\color{cyan}\mdseries-{-}-}}

\newcommand{\hl}[1]{{\color{RoyalBlue}\texttt{#1}}}
\newcommand{\hlopt}[1]{{\color{blue}\texttt{#1}}}
\newcommand{\hlstr}[1]{{\color{BrickRed}\texttt{#1}}}
\newcommand{\many}[1]{\overline{#1}}
\newcommand{\Div}{\ |\ }

\definecolor{mygreen}{rgb}{0,0.6,0}
\definecolor{mygray}{rgb}{0.5,0.5,0.5}
\definecolor{mymauve}{rgb}{0.58,0,0.82}
 
\definecolor{darkgray}{rgb}{.4,.4,.4}
\definecolor{purple}{rgb}{0.65, 0.12, 0.82}
 
\lstdefinelanguage{JavaScript}{
keywords={typeof, new, true, false, catch, function, return, null, catch, switch, var, if, in, while, do, else, case, break, let, for, range, call},
keywordstyle=\color{purple},
ndkeywords={class, export, boolean, throw, implements, import, this},
ndkeywordstyle=\color{darkgray}\bfseries,
identifierstyle=\color{black},
sensitive=false,
comment=[l]{//},
morecomment=[s]{/*}{*/},
commentstyle=\color{mygreen}\ttfamily,
stringstyle=\color{red}\ttfamily,
morestring=[b]',
numbers=left,
numberstyle=\footnotesize,
numbersep=9pt,
morestring=[b]",
backgroundcolor=\color{Gold1!20}
}
 
\lstdefinelanguage{CostExp}{
keywords={eq,in,where,is, max, min},
morekeywords=[2]{minimal, maximal, latency},
keywordstyle=\color{blue},
keywordstyle=[2]\color{green!50!black},
numbersep=9pt,
backgroundcolor=\color{Green!05}
}
 
\newif\ifcomments
\commentstrue

\newcommand{\Keyword}[1]{\ensuremath{\mathtt{#1}}}
\newcommand{\blue}[1]{{\color{RoyalBlue}#1}}
\newcommand{\red}[1]{{\color{RedOrange}#1}}
\newcommand{\purple}[1]{{\color{Purple}#1}}

\newcommand{\Pythonminus}{\textsf{mini Serverless Language}}
\newcommand{\miniSL}{\textsf{miniSL}\xspace}

\newcommand{\Int}{\texttt{Int}}

\newcommand{\semi}{~{\mbox{\tt ;}}~}

\newcommand{\ifte}[3]{\purple{\Keyword{if}}~\mbox{{\tt (}}#1\mbox{{\tt )~\{}}~#2~\mbox{{\tt \}}}~\purple{\Keyword{else}}~\mbox{{\tt \{}}~#3~\mbox{{\tt \}}}}

\newcommand{\for}[3]{\purple{\Keyword{for}}~\mbox{\tt (}#1~\ensuremath{\mathtt{\purple{\Keyword{in}}~\purple{\Keyword{range}}}}\mbox{\tt (0,}#2\mbox{\tt ))\{}~#3~\mbox{\tt \}}}

\let\oldcheckmark\checkmark
\renewcommand{\checkmark}{\blue{\oldcheckmark}}
\let\oldmu\mu
\renewcommand{\mu}{\blue{\oldmu}}

\let\oldsigma\sigma
\renewcommand{\sigma}{\red{\oldsigma}}

\let\oldalpha\alpha
\renewcommand{\alpha}{\red{\oldalpha}}

\newcommand{\C}{{\mathsf{C}}}
\newcommand{\Q}{{\mathsf{Q}}}
\newcommand{\F}{{\mathsf{F}}}
\newcommand{\G}{{\mathsf{G}}}

\newcommand{\Stm}{{\mathsf{S}}}

\newcommand{\E}{{\mathsf{E}}}

\newcommand{\call}{\purple{\texttt{call}}}

\newcommand{\coste}{\mathbb{e}}

\newcommand{\p}{\blue{p}}
\newcommand{\pp}{\blue{p'}}
\newcommand{\pone}{\blue{p\ensuremath{_1}}}
\newcommand{\ptwo}{\blue{p\ensuremath{_2}}}

\newcommand{\iteri}{\blue{i}}
\newcommand{\iterj}{\blue{j}}
\newcommand{\h}{\mathtt{h}}
\newcommand{\g}{\mathtt{g}}

\newcommand{\sjudge}[5]{#1 \vdash^{#2}_{#3} #4 : #5}

\newcommand{\bigfract}[2]{\frac{^{\textstyle #1}}{_{\textstyle #2}}}
\newcommand{\rulename}[1]{{\small {\sc[#1]}}}

\newcommand{\rulenamex}[1]{\mbox{\tiny [{\sc #1}]}}

\def \mathrule #1#2#3{	\begin{array}{l} 
                       	\rulenamex{#1}
                       	\\ 
                      	 \bigfract{#2}{#3}	
                       	\end{array}
					 }

\def \mathax #1#2{\begin{array}{l} 
                  \rulenamex{#1} 
                  \\ 
                  #2
                  \end{array}
                  }

\newcommand{\wt}[1]{\vect{#1}}

\newcommand{\eqdef}{\stackrel{\textsf{\tiny def}}{=}}

\newcommand{\vect}[1]{\overline{#1}}

\newcommand{\var}[1]{\mathit{var}(#1)}

\newcommand{\dom}[1]{\mathit{dom}(#1)}

\let\oldvarepsilon\varepsilon
\renewcommand{\varepsilon}{\blue{\Keyword{\oldvarepsilon}}}

\newcommand{\APP}{\texttt{APP}}
\newcommand{\cAPP}{\texttt{cAPP}}

\let\oldTheta\Theta
\renewcommand{\Theta}{\red{\oldTheta}}

\let\oldLongrightarrow\Longrightarrow
\renewcommand{\Longrightarrow}{\red{\oldLongrightarrow}}

\title{Serverless Scheduling Policies based on Cost Analysis}
\author{Giuseppe De Palma\(^1\),
Saverio Giallorenzo\(^{1,2}\),
Cosimo Laneve\(^{1}\),\\
Jacopo Mauro\(^{3}\),
Matteo Trentin\(^{1,3}\),
Gianluigi Zavattaro\(^{1,2}\)
\institute{\(^{1}\)Universit\`a di Bologna, Italy}
\institute{\(^{2}\)Sophia Antipolis, INRIA, France}
\institute{\(^{3}\)University of Southern Denmark}
}
\def\titlerunning{Serverless Scheduling Policies based on Cost Analysis}
\def\authorrunning{G. De Palma et al.}
\begin{document}
\maketitle

\begin{abstract}
%
Current proprietary and open-source serverless platforms follow opinionated,
hardcoded scheduling policies to deploy the functions to be executed over the
available workers. Such policies may decrease the performance and the security
of the application due to locality issues (e.g., functions executed by workers
far from the databases to be accessed). These limitations are partially overcome
by the adoption of {\APP}, a new platform-agnostic declarative language that
allows serverless platforms to support multiple scheduling logics. Defining the
``right'' scheduling policy in {\APP} is far from being a trivial task since it
often requires rounds of refinement involving knowledge of the underlying
infrastructure, guesswork, and empirical testing.

In this paper, we start investigating how information derived from static
analysis could be incorporated into {\APP} scheduling function policies to help
users select the best-performing workers at function allocation. We substantiate
our proposal by presenting a pipeline able to extract cost equations from
functions' code, synthesising cost expressions through the usage of
off-the-shelf solvers, and extending {\APP} allocation policies to consider
this information.
\end{abstract}

\section{Introduction}

Serverless is a cloud-based service that lets users deploy applications as
compositions of stateless functions, with all system administration tasks
delegated to the platform.
Serverless has two main advantages for users: it saves them time by handling
resource allocation, maintenance, and scaling, and it reduces costs by charging
only for the resources used to perform work since users do not have to pay fur running idle servers~\cite{JSSTKPSMKYGPSP19}.
Several managed serverless offerings are available from popular cloud
providers like Amazon AWS Lambda, Google Cloud Functions, and Microsoft Azure
Functions, as well as open-source alternatives such as OpenWhisk, OpenFaaS,
OpenLambda, and Fission. In all cases, the platform manages the allocation of
function executions across available computing resources or workers, by adopting
platform-dependent policies. However, the execution times of the functions are not independent of the workers since effects like
\emph{data locality} (the latencies to access data depending on the node position)
can increase the run time of functions~\cite{HSHVAA16}.

\begin{figure}[t]
\begin{minipage}{0.40\columnwidth}
\includegraphics[width=1\textwidth]{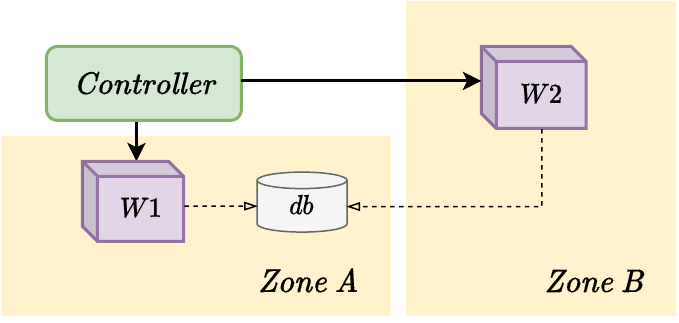} 
\end{minipage}
\hfill
\begin{minipage}{0.35\textwidth}
\vspace{2em}{\small
\begin{lstlisting}[language=yaml,backgroundcolor=\color{Gray!10},mathescape=true]
- db_query:
  - $\hl{workers}$: 
    - $\hl{wrk}$: W1
    - $\hl{wrk}$: W2
    $\hl{strategy}$: $\hlopt{best\_first}$
\end{lstlisting}}
\end{minipage}
\caption{Example of function-execution scheduling problem and \APP{} script.}
\label{fig:example}
\end{figure}

We visualise the issue by commenting on the minimal scenario drawn in
\cref{fig:example}. There, we have two workers, {\tt W1} and {\tt W2}, located
in distinct geographical \emph{Zones} \(A\) and \(B\), respectively. Both
workers can run functions that interact with a database (\emph{db}) located in
\(Zone\ A\). When the function scheduler --- the \emph{Controller} --- receives a
request to execute a function, it must determine which worker to use. To
minimise response time, the function scheduler must take into account the
different computational capabilities of the workers, as well as their current
workloads,
and, for functions that interact with the database, the time to access the
database.
In the example, since {\tt W1} is geographically close to \(db\), it can access
\(db\) with lower latencies than {\tt W2}. 

\APP{}~\cite{PGMZ20,DGMTZ22} is a declarative language recently introduced to
support the \emph{configuration of custom function-execution scheduling
policies}. The \APP{} snippet in \cref{fig:example} codifies the (data) locality
principle of the example. Concretely, in the platform, we associate the
functions that access \(db\) with a tag, called \texttt{db\_query}. Then, we
include the scheduling rule in the snippet to specify that every function
tagged \texttt{db\_query} can run on either {\tt W1} or {\tt W2}, and the
\hl{strategy} to follow when choosing between them is
\mbox{\hlopt{best\_first}}, i.e., select the first worker in top-down order of
appearance (hence giving priority to the worker {\tt W1} if available and not overloaded).

By featuring customised function scheduling policies, \APP{} allows one to
disentangle from platform-dependent allocation rules. This opens the problem of
finding the most appropriate scheduling for serverless applications.
The approach currently adopted by \APP{} is to feature only a few generic
well-established strategies, like the foregoing \mbox{\hlopt{best\_first}}. The
policies are selected \emph{manually}, when the \APP{} script is written, based
on the developer's insights on the behaviour of their functions.
%

In this paper, we propose the adoption of automatic procedures to define
function scheduling policies based on information derived with a static analysis
of the functions. Our approach relies on three main steps: (\emph{i}) the
definition of code analysis techniques for extracting meaningful scheduling
information from function sources; (\emph{ii}) the evaluation of scheduling
information by a(n off-the-shelf) solver that returns cost expressions;
(\emph{iii}) the extension of \APP{} to support allocation strategies
depending on such expressions.
In particular, we discuss the applicability of our approach on a minimal
language for programming functions in serverless applications.

We start in Section~\ref{sec:miniSL} by defining our minimal language 
called {\miniSL} (standing for {\sf mini Serverless Language}) which
includes constructs for specifying computation flow (via \texttt{if} and
\texttt{for} constructs) and for service invocation (via a \texttt{call}
construct).
%
Then, by following~\cite{Laneve2017,Laneve2022}, we describe in
Section~\ref{sec:inference} how to exploit a (behavioural) type system to
automatically extract a set of equations from function source codes that define
meaningful configuration costs. In Section~\ref{sec:inference} we also discuss
how equations can be fed to off-the-shelf cost analyser (e.g., {\tt
PUBS}~\cite{PUBS} or {\tt CoFloCo}~\cite{COFLOCO}) to compute cost expressions
quantifying over-approximations of the considered configuration costs. These
expressions are then used in Section~\ref{sec:extension} to define 
scheduling policies in an extension of {\APP}, dubbed {\cAPP}. Finally, in
Section~\ref{sec:conclusion} we draw some concluding remarks.

%



\section{The \Pythonminus}
\label{sec:miniSL}
The {\Pythonminus}, shortened into {\miniSL}, is a minimal calculus that we use
to define the functions' behaviour in serverless computing. In particular,
{\miniSL} focuses only on core constructs to define operations to access
services, conditional behaviour with simple guards, and iterations.

Function executions are triggered by events. At triggering time, a function
receives a sequence of invocation parameters: for this reason, we assume a countable 
set of \emph{parameter names}, ranged over by $\p$, $\pp$. We also consider a countable set of 
\emph{counters}, ranged over by $\iteri$, $\iterj$, used as indexes in iteration
statements.
%
%
%
Integer numbers are represented by $n$; service names are represented by $\h$, $\g$, 
$\cdots$.
The syntax of {\miniSL} is as follows (we use over-lines to denote 
sequences, e.g., $\pone, \ptwo$ could be an instance of $\overline{\p}$):
%
{\small\[
\begin{array}{r@{\quad}l}
\F\ ::= & \mbox{{\tt (}} \overline{\p} \mbox{{\tt ) => \{}} \; \Stm \; \mbox{{\tt \}}}
\\
\Stm\ ::=  & \varepsilon 
\quad | \quad \call \; \h (\overline{\E}) \;\, \Stm
\quad | \quad \ifte{\G}{\Stm}{\Stm} \quad | \quad \for{\iteri}{\E}{\Stm}
\\
\G\ ::= &  \E \quad | \quad \call \; \h (\overline{\E})
\\
\E\ ::= & n \quad | \quad \iteri \quad | \quad \p 
\quad | \quad  \E ~\sharp~ \E 
\\
\sharp\ ::= & \mbox{{\tt +}} \quad | \quad \mbox{{\tt -}} \quad | \quad 
\mbox{{\tt >}} \quad | \quad \mbox{{\tt ==}} \quad | \quad \mbox{{\tt >=}}  
\quad | \quad \mbox{{\tt \&\&}} \quad | \quad \mbox{{\tt *}} \quad | \quad \mbox{{\tt /}}                 
\end{array}
\]}

A \emph{function} $\F$ associates to a sequence of parameters $\overline{\p}$ 
a statement $\Stm$ which is executed at every occurrence of the triggering event.
%
\emph{Statements} include the empty statement $\varepsilon$ 
(which is always omitted when the statement is not empty); 
calls to external services by means of the $\call$ keyword;
the conditional and iteration statements.
The guard of a conditional statement could be either a boolean
expression or a call to an external service which, in this case, 
is expected to return a boolean value. 
The language supports standard expressions in which it is possible
to use integer numbers and counters. Notice that, in our simple
language, the iteration statement considers an iteration variable
ranging from $0$ to the value of an expression $\E$ evaluated
when the first iteration starts. 


%

In the rest of the paper, we assume all programs to be well-formed
so that all names are correctly used,
i.e., counters are declared before they are used
and when we use $\p$, such $\p$ is an invocation parameter. 
Similarly, for each expression used in the range of
an iteration construct, we assume that its evaluation
generates an integer, and 
for each service invocation $\call \; \h(\overline{E})$, we assume that $\h$ is a correct service name and $\overline{E}$ is a sequence of expressions generating
correct values to be passed to that service. Calls to services include serverless invocations, which possibly execute on a different worker of the caller.

We illustrate {\miniSL} by means of three examples.
As a first example, consider the code in Listing~\ref{lst.if_internal} representing the call of a function that selects a functionality based on the characteristic of the invoker.

\begin{lstlisting}[xleftmargin=.3\textwidth,linewidth=.7\textwidth,language=JavaScript,mathescape=true,label=lst.if_internal,caption={Function with a conditional statement guarded by an expression.}]
( isPremiumUser, par ) => {
  if( isPremiumUser ) {
    call PremiumService( par )
  } else {
    call BasicService( par )
  }
}
\end{lstlisting}
  
\noindent
This code may invoke either a {\ttfamily PremiumService} or a {\ttfamily
BasicService} depending on whether it has been triggered by a premium user or
not. The parameter {\ttfamily isPremiumUser} is a value indicating whether the
user is a premium member (when the value is true) or not (when the value is
false). The other invocation parameter {\ttfamily par} must be forwarded to the
invoked service. For the purposes of this paper, this example is relevant
because if we want to reduce the latency of this function, the best node to
schedule it could be the one that reduces the latency of the invocation of
either the service {\ttfamily PremiumService} or the service {\ttfamily
BasicService}, depending on whether {\ttfamily isPremiumUser} is true or false,
respectively.


Consider now the following function where differently from the previous
version, it is necessary to call an external service to decide whether
we are serving a premium or a basic user.

\begin{center}
\begin{minipage}{\textwidth}
\begin{lstlisting}[xleftmargin=.25\textwidth,linewidth=.75\textwidth,language=JavaScript,mathescape=true,label=lst.if_external,caption={Function with a conditional statement guarded by an invocation to external service.}]
( username, par ) => {
  if( call IsPremiumUser( username ) ) {
    call PremiumService( par )
  } else {
    call BasicService( par )
  }
}
\end{lstlisting}
\end{minipage}
\end{center}

Notice that, in this case, the first parameter carries an attribute of the user
(its name) but it does not indicate (with a boolean value) whether it is a
premium user or not. Instead, the necessary boolean value is returned by the
external service {\ttfamily IsPremiumUser} that checks the username and returns
true only if that username corresponds to that of a premium user. In this case,
it is difficult to predict the best worker to execute such a function, because
the branch that will be selected is not known at function scheduling time. If
the user triggering the event is a premium member, the expected execution time
of the function is the sum of the latencies of the service invocations of
{\ttfamily IsPremiumUser} and {\ttfamily PremiumService} while, if the user is
not a premium member, the expected execution time is the sum of the latencies of
the services {\ttfamily IsPremiumUser} and {\ttfamily BasicService}. As an
(over-)approximation of the expected delay, we could consider the worst execution
time, i.e., the sum of the latency of the service {\ttfamily IsPremiumUser} plus
the maximum between the latencies of the services {\ttfamily PremiumService} and
{\ttfamily BasicService}. At scheduling time, we could select the best worker as
the one giving the best guarantees in the worst case, e.g., the one with the
best over-approximation.


Consider now a function triggering a sequence of map-reduce jobs.

\begin{lstlisting}[xleftmargin=.3\textwidth,linewidth=.7\textwidth,language=JavaScript,mathescape=true,label=lst.forecast,caption={Function implementing a map-reduce logic.}]
( jobs, m, r ) => {
  for(i in range(0, m)) {
    call Map(jobs, i)
    for(j in range(0, r)) {
      call Reduce(jobs, i, j)
    }
  }
}
\end{lstlisting}

The parameter {\ttfamily jobs} describes a sequence of map-reduce jobs. The
number of jobs is indicated by the parameter {\ttfamily m}. The ``map'' phase,
which generates {\ttfamily m} ``reduce'' subtasks, is implemented by an external
service {\ttfamily Map} that receives the {\ttfamily jobs} and the specific
index {\ttfamily i} of the job to be mapped. The ``reduce'' subtasks are
implemented by an external service {\ttfamily Reduce} that receives the
{\ttfamily jobs}, the specific index {\ttfamily i} of the job under execution,
and the specific index {\ttfamily j} of the ``reduce'' subtask to be executed
--- for every {\ttfamily i}, there are {\ttfamily r} such subtasks. In this
case, the expected latency of the entire function is given by the sum of
{\ttfamily m} times the latency of the service {\ttfamily Map} and of {\ttfamily
m} $\times$ {\ttfamily r} times the latency of the service {\ttfamily Reduce}.
Given that such latency could be high, a user could be interested to run the
function on a worker, only if the expected overall latency is below a given
threshold.


\section{The inference of cost expressions}
\label{sec:inference}
In this section, we formalise how one can extract a cost program from \miniSL
code. Once extracted, we can feed this program to off-the-shelf tools, such
as~\cite{COFLOCO,PUBS}, to calculate the cost expression of the related \miniSL
code.

\noindent
Cost programs are lists of \emph{equations} which are terms
\[
f(\vect{x}) \; = \; \coste + \sum_{i \in 0..n} f_i(\vect{\coste_i}) \qquad \qquad [\; \varphi \;]
\]
where variables occurring in the right-hand side and in $\varphi$ are a subset of $\vect{x}$ and
$f$ and $f_i$ are (cost) function symbols. Every function definition has a right-hand side 
consisting of 
\begin{itemize}
%
%
%
%
\item
a \emph{Presburger arithmetic expression} $\coste$ whose syntax is 
\[
\coste \; ::= \qquad x \quad | \quad q \quad | \quad  \coste +\coste \quad | \quad
\coste - \coste \quad | \quad q * \coste \quad | \quad \mathit{max}(\coste_1, \cdots, \coste_k) 
\]
where $x$ is a variable and $q$ is a positive rational number,

\item
a number of \emph{cost function invocations} $f_i(\vect{\coste_i})$ where $\vect{\coste_i}$
are Presburger arithmetic expressions,
%

\item
the \emph{Presburger guard} $\varphi$ is a \emph{linear conjunctive constraint}, \emph{i.e.}, a conjunction of
\emph{constraints} of the
form $\coste_1 \geq \coste_2$ or $\coste_1 = \coste_2$, where 
both $\coste_1$ and $\coste_2$ are Presburger arithmetic expressions. 
\end{itemize}
The intended meaning of an equation 
$f(\vect{x}) \; = \; \coste + \sum_{i \in 0..n} f_i(\vect{\coste_i})\ \ [\; \varphi \;]$
is that the cost of $f$ is given by $\coste$ and the costs of $f_i(\vect{\coste_i})$, when the guard $\varphi$ is true.
Intuitively, $\coste$ quantifies the specific cost of one execution of $f$ without
taking into account invocations of either auxiliary functions or recursive calls.
Such additional cost is quantified by $\sum_{i \in 0..n} f_i(\vect{\coste_i})$.
The \emph{solution of a cost program} is an expression, quantifying the cost of the 
function symbol in the first equation in the list, which is parametric in the formal parameters 
of the function symbol.

For example, 
the following cost program
\[
\begin{array}{lll@{\qquad}l}
f(N, M) & = & M + f(N-1,M) & [N \geq 1]
\\
f(N,M) & = & 0 & [N=0]
\end{array}
\]
defines a function $f$ that is invoked $N+1$ times and each invocation, excluding the last having cost $0$, 
costs $M$. 
The solution of this cost program is the \emph{cost expression} $N \times M$. 

Our technique associates cost programs to {\miniSL} functions by parsing the corresponding 
codes. In particular, we define a set of (inference) rules that gather fragments of 
cost programs that are then combined in a syntax-directed manner. 
As usual with syntax-directed rules, we use \emph{environments} $\Gamma$, $\Gamma'$,
which are maps. In particular,
\begin{itemize}
\item
$\Gamma$ takes a service $\h$ or a parameter name $\p$ and returns a Presburger arithmetics expression, which is usually a variable. 
For example, if $\Gamma(\mathtt{h}) = X$, then $X$ will appear in the cost expressions of 
{\miniSL} functions using $\mathtt{h}$ and will represent the cost for accessing the service.
As regards parameter names $\p$, $\Gamma(\p)$ 
represents values which are known at
function scheduling time,

\item
$\Gamma$ takes counters $\iteri$ and returns the type {\Int}.
\end{itemize}
When we
write $\Gamma + \iteri : {\Int}$, we assume that $\iteri$ does not belong
to the domain of $\Gamma$. Let $\C$ be a sum of cost of function invocations and let
$\Q$ be a list of equations.
Judgments have the shape
\begin{itemize}
\item
$\sjudge{\Gamma}{}{}{ \E }{\coste}$, meaning that the value of the \emph{integer expression} $\E$ in $\Gamma$ is represented by (the Presburger arithmetic expression) $\coste$,
\item $\sjudge{\Gamma}{}{}{ \E }{\varphi}$, meaning that the value of the \emph{boolean expression} $\E$ in $\Gamma$ is represented by (the Presburger guard) $\varphi$,
\item
$\sjudge{\Gamma}{}{}{ \Stm }{\coste \semi \C \semi \Q }$, 
 meaning that the 
cost of $\Stm$ in the environment $\Gamma$ is $\coste + \C$ given a list $\Q$ of equations,
\item
$\sjudge{\Gamma}{}{}{ \F }{\Q }$, 
meaning that the 
cost of a function $\F$ in the environment $\Gamma$ is the list $\Q$ of equations.
\end{itemize}
We use the notation $\var{\coste}$ to address the set of variables occurring in $\coste$, which is extended to
tuples $\var{\coste_1, \cdots , \coste_n}$ with the standard meaning. Similarly 
$\var{\sum_{i \in 0..n} f_i(\vect{\coste_i})}$ is the union of the sets of variables 
$\var{\vect{\coste_0}}, \cdots , \var{\vect{\coste_n}}$. 

The inference rules for {\miniSL} are reported in Figure~\ref{fig.costrules}. They
compute the cost of a program with respect to the calls to
external services (whose cost is recorded in the environment $\Gamma$). Therefore, if a
{\miniSL} expression (or statement) has no service invocation, its cost is 0.
Notice that
in the rule~\rulename{if-exp} we use the guard $[\; \neg \varphi \;]$, to model 
the negation of a linear conjunctive constraint $\varphi$, even if negation is not
permitted in
Presburger arithmetic. Actually, such notation is syntactic sugar defined as follows: 
\begin{itemize}
\item
let $\neg \varphi$ (the \emph{negation} of a Presburger guard 
$\varphi$) be the \emph{list} of Presburger guards
\[
\begin{array}{rl}
\neg (\coste \geq \coste') \; = &  \, \coste' \geq \coste + 1\, 
\\
\neg (\coste = \coste') \; = &  \, \coste \geq \coste'+1 \semi \coste' \geq \coste+1 \, 
\\
\neg (\coste \wedge \coste') \; = & \neg \coste \semi \neg \coste'
\end{array}
\]
where $\semi$ is the list concatenation operator (the list represents a \emph{disjunction of Presburger guards}),
\item
let 
$\neg \varphi = \; \varphi_1 \semi \cdots \semi \varphi_m \;$, where $\varphi_i$ are 
Presburger guards,
then
\[
\Bigl(f(\vect{x}) \; = \; \coste + \sum_{i \in 0..n} f_i(\vect{\coste_i})\Bigr)
\; [\, \neg \varphi \,] \quad \eqdef \quad 
\Bigl\{
f(\vect{x}) \; = \; \coste + \sum_{i \in 0..n} f_i(\vect{\coste_i}) \quad [\, \varphi_j  \,] \quad | \quad j \in 1..m   \Bigr\} \; .
 \]
\end{itemize}
%
%
%
%
%
%

\begin{figure}[t]
\[
\begin{array}{c}
\mathax{eps}{
	\sjudge{\Gamma}{}{}{\varepsilon}{0 \semi \emptyset \semi \emptyset}
	}
\qquad
\mathrule{call}{
	\begin{array}{c}
	\Gamma(\h) = \coste
	\qquad
	\sjudge{\Gamma}{}{}{ \Stm }{\coste' \semi \C \semi \Q }
	\end{array}
	}{
	\sjudge{\Gamma}{}{}{\call \; \h(\vect{\E}) \; \; \Stm}{\coste+\coste' \semi \C \semi \Q}
	}
\\
\mathrule{if-exp}{
	\begin{array}{c}
	\sjudge{\Gamma}{}{}{ \E }{\varphi}
	\qquad
	\sjudge{\Gamma}{}{}{ \Stm }{\coste' \semi \C \semi \Q }
	\qquad
	\sjudge{\Gamma}{}{}{ \Stm' }{\coste'' \semi \C' \semi \Q' }
	\qquad \mathit{if}_\ell \; \; \mathit{fresh}
	\\
	\wt{w} = \var{\coste,\coste',\coste''} \cup\var{\C,\C'}
	\qquad
	\Q'' = \left[ 
				\begin{array}{ll}
			     \mathit{if}_\ell(\wt{w}) = \coste' +
			     \C  & [~\varphi~]
		  		 \\ 
				 \mathit{if}_\ell(\wt{w}) = \coste'' + \C' & [\neg \varphi]
			     \end{array} 
		 \right] 
	\end{array}
	}{
	\sjudge{\Gamma}{}{}{ \ifte{\E}{\Stm}{\Stm'} }{0 \semi
	\mathit{if}_\ell(\wt{w}) \semi \Q , \; \Q', \Q'' }
	}
\\
\mathrule{if-call}{
	\begin{array}{c}
	\Gamma(\h) = \coste
	\qquad
	\sjudge{\Gamma}{}{}{ \Stm }{\coste' \semi \C \semi \Q }
	\qquad
	\sjudge{\Gamma}{}{}{ \Stm' }{\coste'' \semi \C' \semi \Q' }
	\end{array}
	}{
	\sjudge{\Gamma}{}{}{ \ifte{ \call \; \h(\vect{\E})}{\Stm}{\Stm'} }{\coste + \mathit{max}(\coste', \coste'') \semi
	\C  + \C' \semi \Q , \; \Q' }
	}
\\
\mathrule{for}{
	\begin{array}{c}
	\sjudge{\Gamma}{}{}{ \E }{\coste}
	\qquad
	\sjudge{\Gamma + \iteri : \Int}{}{}{ \Stm }{\coste' \semi \C \semi \Q }
	\qquad \wt{w} = (\var{\coste,\coste'} \cup \var{\C}) \setminus \iteri 
	\\
	\mathit{for}_\ell \; \; \mathit{fresh}\qquad 
	\Q' = \left[ 
				\begin{array}{ll}
			     \mathit{for}_\ell(\iteri, \wt{w}) = \coste' +
			     \C + \mathit{for}_\ell(\iteri+1, \wt{w}) \quad & [~\coste ~\geq~ \iteri~]
		  		 \\ 
				 \mathit{for}_\ell(\iteri, \; \wt{w}) = 0 & [~\iteri ~\geq~ \coste+1  ~]
			     \end{array} 
		 \right]   
	\end{array}
	}{
	\sjudge{\Gamma}{}{}{ \for{\iteri}{\E}{\Stm} }{0 \semi
	\mathit{for}_\ell(0, \; \wt{w}) \semi  \Q, \; \Q'   }
	}
\\
\mathrule{prg}{
	\begin{array}{c}
	\sjudge{\Gamma}{}{}{ \Stm }{\coste \semi \C \semi \Q }
	\qquad \wt{w} = \var{\vect{\p},\coste} \cup \var{\C}
	\\
	\mathit{main}\; \; \mathit{fresh}\qquad \Q' = \mathit{main}(\wt{w}) = \coste + \C \quad [~]
	\end{array}
	}{
	\sjudge{\Gamma}{}{}{ 
	\mbox{{\tt (}} \vect{\p} \mbox{{\tt ) => \{}} \; \Stm \; \mbox{{\tt \}}} }{\Q', \; \Q   }
}
\end{array}
\]
\caption{\label{fig.costrules} The rules for deriving cost expressions}
\end{figure}
We now comment on the inference rules reported in Figure~\ref{fig.costrules}.\footnote{We omit rules for expressions $\E$ since they are
straightforward: they simply return $\E$ if $\E$ is in Presburger arithmetics.}

Rule~\rulename{call} manages invocation of services: the cost of $\call \; \h(E) \; 
\; \Stm$ is the cost of $\Stm$ plus the cost for accessing the service $\h$.

Rule~\rulename{if-exp} defines 
the cost of conditionals when the guard is a Presburger arithmetic expression
that can be evaluated at function scheduling time. 
We use a corresponding
cost function, $\mathit{if}_\ell$, whose name is fresh,\footnote{We assume that conditionals have pairwise different line-codes and
$\ell$ represents the line-code of the if in the source code.}
to indicate that
the cost of the entire conditional statement is either the cost
of the then-branch or the else-branch, depending on whether the guard is
true or false.
%
As discussed above, the use of the guard $\neg \varphi$ 
generates a list of equations.


Rule~\rulename{if-call} defines an upper bound of the 
cost of conditionals when the guard is an invocation to a service. 
At scheduling time it is not possible to determine whether the guard 
is true or false -- \emph{c.f.}~the second example in Section~\ref{sec:miniSL}.
Therefore the cost of a conditional is the maximum between the cost $\coste' + \C$ of
the then-branch
and the one $\coste'' + \C' $ of the else-branch, plus the cost $\coste$ to 
access to the service  in the guard. However, considering that the expression
$\mathit{max}(\coste + \C, \coste'+ \C')$ is not a valid right-hand side for the equations
in our cost programs, 
we take as over-approximation the expression $\mathit{max}(\coste, \coste') + \C+ \C'$.

As regards iterations, according to~\rulename{for}, its cost is the invocation 
of the corresponding function, $\mathit{for}_\ell$, whose name is fresh (we assume that iterations have pairwise different line-codes). The rule adds the
counter $\iteri$ to $\Gamma$ (please recall that $\Gamma + \iteri : \Int$ entails that
$\iteri \notin \dom{\Gamma}$).
In particular,  the counter $\iteri$ is the first formal parameter of $\mathit{for}_\ell$;
the other parameters are all the variables
in $\coste$, in notation $\var{\coste}$ plus those in the invocations $\C$ 
(minus the $\iteri$). There are two equations for every iteration: one is the case when 
$\iteri$ is out-of-range, hence the cost is $0$, the other is when it is in range and the
cost is the one of the body \emph{plus} the cost of the recursive invocation of 
$\mathit{for}_\ell$ with $\iteri$ increased by 1.

The cost of a {\miniSL} program is defined by~\rulename{prg}. This rule defines
an equation for the function $\mathit{main}$ and puts this equation as the first
one in the list of equations.

As an example, in the following, we apply the rules of Figure~\ref{fig.costrules} to the codes in
Listings~\ref{lst.if_internal},~\ref{lst.if_external} and~\ref{lst.forecast}. 
Let $\Gamma(\mathtt{isPremiumUser}) =u$, $\Gamma(\mathtt{PremiumService}) = P$
and  $\Gamma(\mathtt{BasicService}) = B$.
For Listing~\ref{lst.if_internal} we obtain the cost program
\[
\begin{array}{rl@{\qquad}l}
\mathit{main}(u,P,B) = & \mathit{if}_2(u,P,B) & [\;]
\\
\mathit{if}_2(u,P,B) = & P & [~u = 1~]
\\
\mathit{if}_2(u,P,B) = & B & [~u = 0~]
\end{array}
\]

\noindent
For Listing~\ref{lst.if_external}, let $\Gamma(\mathtt{IsPremiumUser}) = K$.
Then the rules of Figure~\ref{fig.costrules} return 
the single equation
\[
\begin{array}{rl@{\qquad}l}
\mathit{main}(K,P,B) = & K + \mathit{max}(P,B) & [\;]
\end{array}
\]
For ~\ref{lst.forecast}, when $\Gamma(\mathtt{m}) = m$, $\Gamma(\mathtt{r}) = r$, $\Gamma(\mathtt{Map}) = M$
and $\Gamma(\mathtt{Reduce}) = R$, 
the cost program is
\[
\begin{array}{rl@{\qquad}l}
\mathit{main}(m,r,M,R) = & \mathit{for}_2(0,m,r,M,R) & [\;]
\\
\mathit{for}_2(i,m,r,M,R) = & M + \mathit{for}_4(0,r,R) + \mathit{for}_2(i+1,m,r,M,R)
& [~m \geq i~]
\\
\mathit{for}_2(i,m,r,M,R) = & 0 & [~i \geq m+1~]
\\
\mathit{for}_4(j,r,R) = & R + \mathit{for}_4(j+1,r,R)
& [~r \geq j~]
\\
\mathit{for}_4(j,r,R) = & 0 & [~j \geq r+1~]
\end{array}
\]

\noindent
The foregoing cost programs can be fed to
automatic solvers
such as Pubs~\cite{PUBS} and
CoFloCo~\cite{COFLOCO}. 
The evaluation of the cost program for Listing~\ref{lst.if_internal} returns 
$\textit{max}(P,B)$ because $u$ is unknown. On the contrary, if $u$ is known,
it is possible to obtain a more precise evaluation from the solver:
if $u=1$ it is possible to ask the solver to consider $\textit{main}(1,P,B)$
and the solution will be $P$, while if $u=0$ it is possible to ask the solver 
to consider $\textit{main}(0,P,B)$ and the solution will be $B$.
%
The evaluation of $\textit{main}(K,P,B)$ for Listing~\ref{lst.if_external}
gives the expression $ K + \mathit{max}(P,B)$, which is exactly what is written in the equation. This 
is reasonable because, statically, we are not aware of the value returned by the 
invocation of {\tt IsPremiumService}. Last, the evaluation of  the cost program  for 
Listing~\ref{lst.forecast} returns the expression $m \times (M + r \times R)$.

\section{From {\APP} to {\cAPP}}
\label{sec:extension}

We now discuss the extension of \APP{} that we plan to realise, where function
scheduling policies could depend on the costs associated with the possible
execution of the functions on the available workers.

Before discussing the extensions towards \cAPP{}, we briefly introduce the
\APP{} syntax and constructs, reported in Figure~\ref{fig:app_syntax}, as found
in its first incarnation by De Palma et al.~\cite{PGMZ20}

\subsection*{The \APP{} Language}

An \APP{} script is a collection of tagged scheduling policies. The main,
mandatory component of any policy (identified by a \textit{policy\_tag}) are the
\hl{workers} therein, i.e., a collection of labels that identify on which
workers the scheduler can allocate the function. The assumption is that
the environment running \APP{} establishes a 1-to-1 association so that each
worker has a unique, identifying label. A policy, associate to every function a list
of one or more \textit{block}s, each including the \hl{worker} clause to state on which workers the function can be scheduled and two optional parameters: the scheduling \hl{strategy}, followed to select
one of the workers of the block, and an \hl{invalidate} condition, which
determines when a worker cannot host a function. When a selected worker is
invalid, the scheduler tries to allocate the function on the rest of the
available workers in the block. If none of the workers of a block is available,
then the next block is tried. The last clause, \hl{followup}, encompasses a
whole policy and defines what to do when no \textit{block}s of the policy
managed to allocate the function. When set to \hlopt{fail}, the scheduling of
the function fails; when set to \hlopt{default}, the scheduling continues by
following the (special) \texttt{default} policy.

As far as the \hl{strategy} is concerned, it allows the following values:
\hlopt{platform} that applies the default selection strategy of the serverless platform; \hlopt{random} that allocates functions stochastically among the workers
of the block following a uniform distribution; \hlopt{best-first} that allocates
functions on workers based on their top-down order of appearance in the block.
The options for the \hl{invalidate} are instead: \hlopt{overload} that invalidates a worker based on the default invalidation
control of the platform; \hlopt{capacity\_used} that invalidates a worker if it uses more than a given percentage
threshold of memory;
\mbox{\hlopt{max\_concurrent\_invocations}} that invalidates a worker if a given number of function invocations are already currently executed on the worker.

\begin{figure}[t]
    \[
        \begin{array}{lll}
            \textit{policy\_tag} & \in       & \textit{Identifiers} \ \cup \ \{ \texttt{default} \}
            \qquad \textit{worker\_label} \in \textit{Identifiers} \qquad n \ \in \ \mathbb{N}

            \\[.5em]
            \textit{app}         & \Coloneqq & \many{\textit{tag}}
            \\[.5em]
            \textit{tag}         & \Coloneqq & \textit{policy\_tag} \ \hlstr{:} \ \many{\texttt{-}\ \textit{block} } \ \textit{followup}?
            \\[.5em]
            \textit{block}       & \Coloneqq & \hl{workers} \hlstr{:} \ \ [\ \hlopt{*} \Div
            \many{\texttt{-}\ \textit{worker\_label}} \ ]
            \\
                                 &           & (\hl{strategy}\hlstr{:} \ \ [\
            \hlopt{random}
            \Div \hlopt{platform}
            \Div \hlopt{best\_first}
            \ ])?
            \\
                                 &           & (\hl{invalidate} \hlstr{:} \ [\
            \hlopt{capacity\_used} \ \hlstr{:} \ n \hlopt{\%}
            \\ &&  \hspace{2.4cm} \Div \hlopt{max\_concurrent\_invocations}\hlstr{:} \ n
            \\ &&  \hspace{2.4cm} \Div \hlopt{overload}
            \ ])?
            \\[.5em]
            \textit{followup}    & \Coloneqq &
            \hl{followup}\hlstr{:} \ [\
            \hlopt{default}
            \Div \hlopt{fail}
            \ ]
        \end{array}
    \]
    \caption{\label{fig:app_syntax}The \APP{} syntax.\vspace{-1em}}
\end{figure}

\subsection*{Towards \cAPP{}}

Our proposal to extend \APP{} to handle cost-aware scheduling policies entails
two major modifications: (\emph{i}) extending the \APP{} language to express
cost-aware scheduling policies, (\emph{ii}) implementing a new controller that
selects the correct worker following the cost-aware policies.

As far as (i) is concerned, we discuss at least two relevant ways in which costs
can be used. The first one is a new selection strategy named
\hlopt{min\_latency}. Such a strategy selects, among some available workers, the
one which minimises a given cost expression. The second one is a new
invalidation condition named \hlopt{max\_latency}. Such a condition
invalidates a worker in case the corresponding cost expression is greater
than a given threshold.

We dub {\cAPP} the cost-aware extension of {\APP} and illustrate its main features by showing examples of \cAPP{} scripts that target the functions in Listings \ref{lst.if_internal}--\ref{lst.forecast}.

\noindent\hfil\begin{minipage}{.41\textwidth}
{\small
\begin{lstlisting}[linewidth=\textwidth,language=yaml,backgroundcolor=\color{Gray!10},mathescape=true,label=lst.app_premUser,caption= {\cAPP} script for Listings \ref{lst.if_internal} and \ref{lst.if_external}.]
- premUser:
 - $\hl{workers}$: 
     - $\hl{wrk}$: W1
     - $\hl{wrk}$: W2
   $\hl{strategy}$: $\hlopt{min\_latency}$
\end{lstlisting}
}
\end{minipage}
\hfil
\begin{minipage}{.41\textwidth}
{\small
\begin{lstlisting}[linewidth=\textwidth,language=yaml,backgroundcolor=\color{Gray!10},mathescape=true,label=lst.app_mapReduce,caption= {\cAPP} script for Listing \ref{lst.forecast}.]
- mapReduce :
  - $\hl{workers}$:
    - $\hl{wrk}$: W1
    - $\hl{wrk}$: W2
  $\hl{strategy}$: $\hlopt{random}$
  $\hl{invalidate}$: 
    $\hlopt{max_latency}$: 300
\end{lstlisting}
}
\end{minipage}

In Listing \ref{lst.app_premUser}, we define a \cAPP{} script where we assume to
associate the tag \texttt{premUser} to both the functions at Listing
\ref{lst.if_internal} and \ref{lst.if_external}. In the script, we specify to
follow the logic \hlopt{min\_latency} to select among the two workers,
\texttt{W1} and \texttt{W2} listed in the \hl{workers} clause,
and prioritises the one for which the solution of the cost expression is minimal.
%
%

\begin{figure}
\begin{adjustbox}{width=\textwidth}
\input{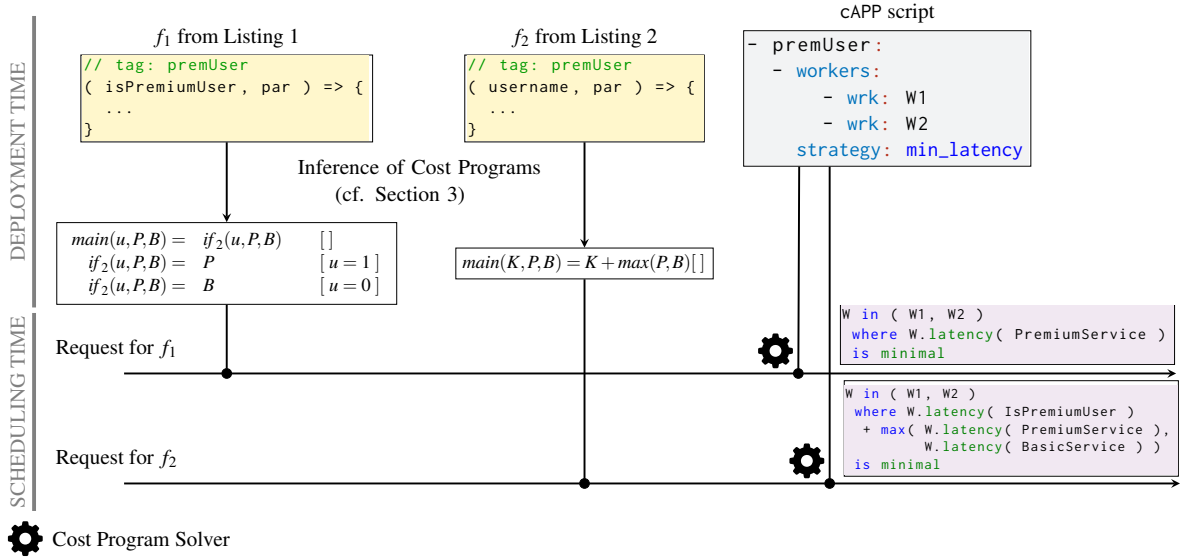}
\end{adjustbox}
\caption{Flow followed, from deployment to scheduling, of the functions at Listings \ref{lst.if_internal} and \ref{lst.if_external}.}
\label{fig:cappScheduling}
\end{figure}

To better illustrate the phases of the \hlopt{min\_latency} strategy, we depict
in \cref{fig:cappScheduling} the flow, from the deployment of the \cAPP{} script to the scheduling of the functions in Listings \ref{lst.if_internal} and \ref{lst.if_external}. When the {\cAPP} script is created, the association between the functions code and
their {\cAPP} script is specified by tagging the two functions
with  
 \lstinline[language=javascript,
 basicstyle=\normalsize\ttfamily]|//tag:premUser|.
In this phase, assuming the scheduling policy of the \cAPP{} script requires the computation of the functions cost, the code of the functions is used to infer the corresponding cost programs.
When the functions are invoked, i.e., at scheduling time, we can
compute the solution of the cost program, given the knowledge of the invocation
parameters. For instance, for the function in Listings \ref{lst.if_internal},
 it is possible to invoke the solver with either $\textit{main}(1,P,B)$ or 
$\textit{main}(0,P,B)$ depending on the actual invocation parameter.
\cref{fig:cappScheduling} illustrates this last part with
the horizontal ``request'' lines found at the bottom. In particular, when we
receive a request for the function at Listing \ref{lst.if_internal}, we take its
cost program (represented by the intersection point on the left) and its
corresponding \cAPP{} policy to implement the expected scheduling policy.
We can implement this behaviour in two steps. First,  the solver solves the cost programs (depicted by the gear); then, we compute the obtained cost expression for each of the possible workers (in this case, \texttt{W1} and \texttt{W2}) by instantiating the parameter representing the cost of invocation of the external services, with an estimation of the latencies from the considered workers.
In this case, given the \hlopt{min\_latency} strategy, the worker that
minimises the 
\lstinline[language=CostExp,basicstyle=\normalsize\ttfamily]|latency| 
to contact \texttt{PremiumService} will be selected.
%
This last step regards the second point (\emph{ii}) mentioned at the beginning of this
section, i.e., the modifications we need to perform on the controller to let it
execute the newly introduced cost-aware strategies at scheduling time. 

\begin{figure}
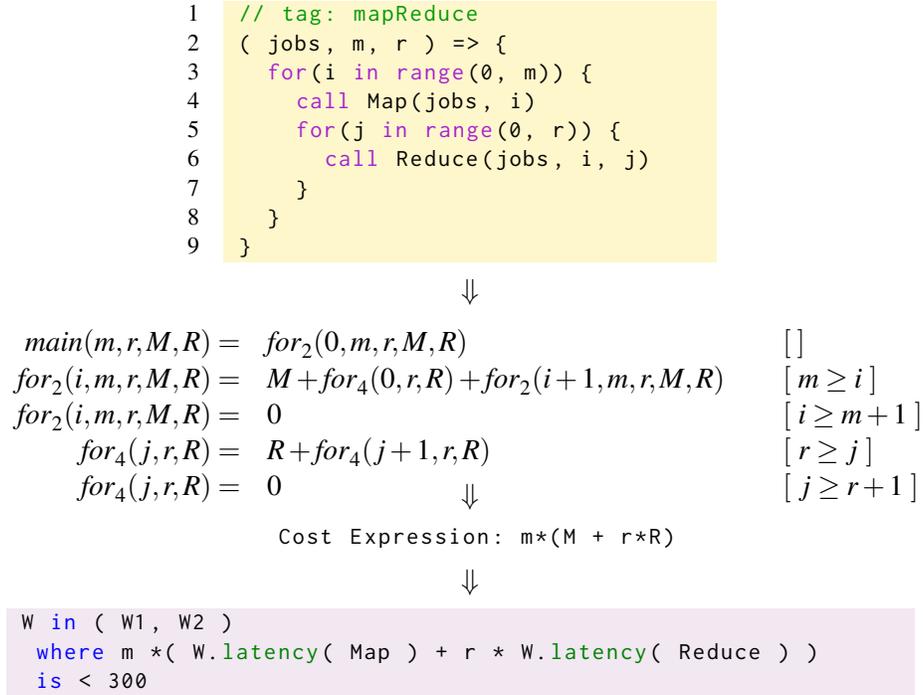

 \begin{center}
 \begin{minipage}{.82\columnwidth}
 \begin{lstlisting}[xleftmargin=.25\textwidth,linewidth=.75\textwidth,language=JavaScript,basicstyle={\ttfamily\footnotesize}, backgroundcolor=\color{Gold1!20},mathescape=true]
 // tag: mapReduce
 ( jobs, m, r ) => {
   for(i in range(0, m)) {
     call Map(jobs, i)
     for(j in range(0, r)) {
       call Reduce(jobs, i, j)
     }
   }
 }
\end{lstlisting}
\vspace{-2em}\[\Downarrow\]\vspace{-1em}
\[
\begin{array}{rl@{\qquad}l}
\mathit{main}(m,r,M,R) = & \mathit{for}_2(0,m,r,M,R) & [\;]
\\
\mathit{for}_2(i,m,r,M,R) = & M + \mathit{for}_4(0,r,R) + \mathit{for}_2(i+1,m,r,M,R)
& [~m \geq i~]
\\
\mathit{for}_2(i,m,r,M,R) = & 0 & [~i \geq m+1~]
\\
\mathit{for}_4(j,r,R) = & R + \mathit{for}_4(j+1,r,R)
& [~r \geq j~]
\\
\mathit{for}_4(j,r,R) = & 0 & [~j \geq r+1~]
\end{array}
\]
\vspace{-2em}\[\Downarrow\]\vspace{-2em}
\begin{lstlisting}[xleftmargin=.29\textwidth,linewidth=.71\textwidth,language=Java,mathescape=true]
 Cost Expression: m*(M + r*R)
\end{lstlisting}
\vspace{-2em}\[\Downarrow\]\vspace{-2em}
\begin{lstlisting}[xleftmargin=.03\textwidth,linewidth=.95\textwidth,language=CostExp,backgroundcolor=\color{Purple!10},mathescape=true]
 W in ( W1, W2 ) 
  where m *( W.latency( Map ) + r * W.latency( Reduce ) )
  is < 300
\end{lstlisting}
\end{minipage}
\end{center}
\vspace{-2em}
\caption{\label{fig:cappMapReduce}The map-reduce function, its cost analysis, and scheduling invalidation logic.}
\end{figure}

For \hlopt{max\_latency}, once a worker is selected using a given strategy, its
corresponding cost is computed in order to check whether the selection is
invalid (i.e., if we can consider the worker able to execute the function, given
the invalidation constraints of the script). To illustrate this second
occurrence, we look at the {\cAPP} code we wrote for the map-reduce function in
Listing \ref{lst.app_mapReduce}, and we illustrate it using
\cref{fig:cappMapReduce}. As seen above, we start (top-most box) from the
deployment phase, where we tag the function
(\lstinline[language=javascript,basicstyle=\normalsize\ttfamily]|//tag:mapReduce|)
and we proceed to compute its cost program, obtaining the associated cost
expression. Then, when we receive a request for that function, we trigger the
execution of the \cAPP{} policy, which selects one of the two workers
\texttt{W1} or \texttt{W2} at \hlopt{random} and checks their validity following
the logic shown at the bottom of \cref{fig:cappMapReduce}, i.e., 
we solve the cost program and then compute the corresponding cost expression by
replacing the parameters \texttt{m} and \texttt{r} with the  
\lstinline[language=CostExp,basicstyle=\normalsize\ttfamily]|latency| to contact
the \texttt{Map} and \texttt{Reduce} services from the selected worker, 
and possibly invalidate it if the computed value is greater than \texttt{300}.
In the function's code, for simplicity, we abstract away the coordination
logic between \texttt{Map} and \texttt{Reduce} (which usually performs a multipoint scatter-gather behaviour) by offloading it to external services (e.g., a database contacted by the functions).

These new \hl{strategy} and \hl{invalidate} parameters added for \cAPP{} interact with the cost-inference logic presented in Section~\ref{sec:inference}. As shown in Figure~\ref{fig:cappScheduling}, the definition of the \hl{strategy} and \hl{invalidate} parameters, as well as the cost inference, happen independently, when the \cAPP{} script is deployed. A strategy indeed (e.g., \hlopt{min\_latency}) is not tied to any specific cost expression. For example, the user can define the \texttt{premUser} policy (see the \cAPP{} script on the right-hand side of Figure~\ref{fig:cappScheduling}) before having deployed any function with that tag.
When functions are deployed on the platform (centre and left-hand side of Figure~\ref{fig:cappScheduling}), the \cAPP{} runtime performs the inference of programs' costs. When instead a request for the execution of a function reaches the platform, the \cAPP{} use the cost expressions and create the logic of selection/invalidation down to its runtime form.
For instance, in Figure~\ref{fig:cappScheduling}, the scheduling of function \(f_1\) compiles the \hlopt{min\_latency} logic using the reduced form \(P\) (the cost of accessing service \texttt{PremiumService}) since at scheduling time the parameter \texttt{isPremiumUser} (represented by the variable \(u\) in the related cost equations in Figure~\ref{fig:cappScheduling}) is known, which in the example we value to 1 (i.e., the request is from a premium user). From the reduced cost expression we can obtain the \cAPP{} selection logic on the right-hand side of Figure~\ref{fig:cappScheduling}: select that worker, among the one provided in the \cAPP{} block, that minimises (\lstinline[basicstyle={\ttfamily\normalsize},language=CostExp]{is minimal}) the \lstinline[basicstyle={\ttfamily\normalsize},language=CostExp]{latency} of interaction with the \lstinline[basicstyle={\ttfamily\normalsize},language=CostExp]{PremiumService} service.

For completeness, we can draw a parallel example for the invalidation parameter by looking at Figure~\ref{fig:cappMapReduce}. There, once we have a request for the map-reduce function, we take the cost expression calculated at deployment time, whose (integer) values represented by \texttt{m} are \texttt{r} are known at scheduling time, and we compile the \hl{invalidate} logic, \hlopt{max\_latency}\texttt{:300} --- for the map-reduce function, the logic declares invalid any worker whose cost \lstinline[basicstyle={\ttfamily\normalsize},language=CostExp]{m *( W.latency( Map ) + r * W.latency( Reduce ) )} exceeds the set \texttt{300} threshold.

\section{Conclusion}
\label{sec:conclusion}

We have presented a proposal for an extension of the {\APP} language, called
{\cAPP}, to make function scheduling cost-aware. Concretely, the extension adds
new syntactic fragments to {\APP} so that programmers can govern the scheduling
of functions towards those execution nodes that minimise their calculated
latency (e.g., increasing serverless function performance) and avoids running
functions on nodes whose execution time would exceed a maximal response time
defined by the user (e.g., enforcing quality-of-service constraints). The main
technical insights behind the extension include the usage of inference rules to
extract cost equations from the source code of the deployed functions and
exploiting dedicated solvers to compute the cost of executing a function, given
its code and input parameters.

Growing our proposal into a usable {\APP} extension is manyfold.
The cost inference of \cref{sec:miniSL} programs is under active development at the time of writing.\footnote{\url{https://github.com/minosse99/CostCompiler}} While the solution of the cost equations can be done by off-the-shelf tools (e.g.,CoFloCo~\cite{COFLOCO}), another important component to develop is the {\cAPP} runtime to
%
generate {\cAPP} rules from the cost equations when functions are scheduled and
interact with the workers available in the
platform to collect the measures that characterise the costs sustained by the
workers (e.g., the latency endured by a worker when contacting a given service).

Implementing the {\cAPP} runtime and proving the feasibility of cost-aware function scheduling is only the first move
along the way. Indeed, in \cref{sec:extension} (illustrated in
\cref{fig:cappScheduling}) we described a na\"ive approach where we solve the
cost equations of an invoked function at scheduling time, but this computation
step could delay the scheduling of the function.
This challenge calls for further investigation. On the one hand, we shall
investigate if the problem presents itself in practice, i.e., if developers
would actually write functions whose cost equations take too much time for the
available engines to solve. On the other hand, we envision working on models and
techniques that can make the problem treatable (e.g., via heuristics and
over-approximations), possibly complementing the former with architectural
solutions, like the inclusion of caching systems that allows us to compute the
actual cost of function invocations once and timeouts paired with sensible
default strategies which would keep the system responsive.

\section*{Acknowledgement}
Research partly supported by the H2020-MSCA-RISE project ID 778233 ``Behavioural Application Program Interfaces (BEHAPI)'' and by the SERICS project (PE00000014) under the MUR National Recovery and Resilience Plan funded by the European Union - NextGenerationEU.

\bibliographystyle{eptcs}
\bibliography{biblio}
\end{document}

